\documentclass{article}
\usepackage{graphicx}
\begin{document}

\title{Evolutions of Neutron Stars and their Magnetic Fields.}
\author{G.S. Bisnovatyi-Kogan\thanks{
Institute of Space Research, Profsoyuznaya 84/32, 117997 Moscow, Russia;
Email: gkogan@iki.rssi.ru}}
\date{}
\maketitle
\begin{abstract}
Estimations of magnetic fields of neutron
stars, observed as radio and X-ray pulsars, are discussed.
It is shown, that theoretical and observational values for different
types of radiopulsars are in good correspondence.
Radiopulsars in close binaries and millisecond pulsars,
which have passed the stage of disk
accretion (recycled radiopulsars),
have magnetic fields 2-4 orders of magnitude smaller than
ordinary single pulsars. Most probably, the magnetic field
of the neutron star was screened by the infalling material.
Several screening models are considered.
Formation of single recycled pulsars loosing its companion is
discussed. Magnetic fields of some
X-ray pulsars are estimated from the cyclotron line energy. In the case of
Her X-1 this estimation exceeds considerably the value of its magnetic field
obtained from long term observational data related to the beam structure
evolution. Another interpretation of the cyclotron feature, based on the
relativistic dipole radiation mechanism, could remove this discrepancy.
Observational data about soft gamma repeators and their interpretation as
magnetars are critically analyzed.

\end{abstract}

\section{Introduction}

First theoretical estimations of neutron star magnetic fields have been
obtained from the condition of the magnetic flux conservation during
contraction of a normal star to a neutron star \cite{gi64}:

\begin{equation}
\label{ref1}
B_{ns}=B_s\left(\frac{R_s}{R_{ns}}\right)^2.
\end{equation}
For main sequence
stellar magnetic field $B_s=10 \div 100$ Gs, and stellar radius
$R_s=(3 \div 10)R_{\odot}\approx (2 \div 7)10^{11}$ cm, we get
$B_{ns}\approx 4\times 10^{11} \div 5\times 10^{13}$ Gs, for a neutron star
radius $R_{ns}=10^6$ cm. As shown below, this simple estimation occurs to
be in a good correspondence with most observational data.

Neutron stars are born with magnetic fields comparable with the fields of
youngest pulsars. The youngest and the best studied Crab pulsar PSR 0531+21
is about 1000 years old and has a magnetic field about $3.5\cdot 10^{12}$ Gs.
The pulsar in Vela PSR 0833-45 is 10 times older, but has almost the same
magnetic field. Magnetic fields of some oldest pulsars PSR 0826-34 ($3\cdot 10^7$
years old) and PSR 1819-22 ($3\cdot 10^7$ years old) are equal to $1.4\cdot 10^{12}$
and $1.1\cdot 10^{12}$ Gs respectively \cite{mt81}. The simple comparison of these
data gives the indication to low, or may be negligible, damping of the magnetic
field due to Ohmic losses inside the neutron star matter. This conclusion seems to be
plausible, because in very dense strongly degenerate layers of the neutron star, where
the electrical current, producing magnetic fields could flow, the electrical conductivity
is enormously high, so in this case we may expect practically no Ohmic damping. These
simple estimations had been confirmed by the detailed statistical analysis of the big
sample of single radiopulsars \cite{verb},
where the conclusion about low (or negligible) Ohmic
damping of magnetic fields of radiopulsars was obtained quite reliably.

\section{Radio pulsars}
It is now commonly accepted that radio pulsars are rotating
neutron stars with inclined magnetic axes. For rotating with angular
velocity $\Omega$ dipole,
the rotational energy losses \cite{ll62,pa67}
 are determined
as $\dot E= A B^2 \Omega^4$. The rotational energy is
$E=\frac{1}{2} I\Omega^2$,
$I$ is an inertia momentum of the neutron star. By measuring of
$P=2\pi/\Omega$ and $\dot P$, we obtain the observational estimation
of the neutron star magnetic field as

\begin{equation}
\label{ref2}
B^2=\frac{IP\dot P}{4\pi^2 A}.
\end{equation}
It was shown by \cite{gj69} that energy losses by
a pulsar relativistic wind are important also when the magnetic and
rotational axes coincide, and one may use (\ref{ref2}) with
$A=\frac{R_{ns}^6}{6c^3}$, and $B_{ns}$ corresponding to the magnetic pole,
at any inclination angle. Magnetic fields of radio pulsars estimated
using (\ref{ref2}) with observational values of $P$ and $\dot P$ lay in a
wide region between $10^8$ and $10^{13}$ Gs \cite{lgs98}.
There are two distinctly different groups: single radiopulsars with
periods exceeding 0.033 sec, and magnetic fields between $10^{11}$ and
$10^{13}$ Gs, and recycled pulsars (RP), present or former members of
close binary systems with millisecond periods and low magnetic fields between
$10^8$ and $10^{10}$ Gs. Low magnetic field of recycled pulsars is
probably a result of its damping during preceding
accretion stage \cite{bisk74,bisk76}.

\section{Recycled radiopulsars}

During several years after discovery of pulsars only single
objects had been found. It was an impression, that pulsars avoid binaries.
When half (or even more) stars are binaries, this phenomena was explained
either
by pair disruption during supernova explosion leading to pulsar formation,
or by absence of SN explosions at the end of evolution of stars in
close binaries \cite{tr}.

Detailed analysis of the fate of close X-ray binaries in low-mass systems,
namely Her X-1, was done in \cite{bisk74}, and the
conclusion was made that evolution of such system will be ended by formation of
a non-accreting neutron star in close binary, which should become a radio pulsar.
It was shown, that neutron star rotation is accelerated
during disk accretion stage, so the second time born
(recycled) radio pulsar should become visible, provided it has a magnetic field
similar to  other single pulsars. Absence of radio pulsars in close binaries,
in spite of intensive searches could be explained by the only reason: during
accretion stage the magnetic field of the neutron star is screened by the
inflowing accreting gas, so the recycled pulsar should have
$B\sim 10^8--10^{10}$ Gs, 2-4 orders of magnitude smaller than average field
strength of radio pulsars. Discovery of the first binary pulsar \cite{ht75},
and subsequent discovery of more than 50 recycled pulsars
\cite{lor03,lyn04} had confirmed this conclusion: all recycled pulsars have
small values of magnetic fields, as was predicted in \cite{bisk74}.

A simple estimation of the magnetic field screening during accretion
had shown a large potential possibility of such process.
It was estimated in \cite{bk79} that in absence of
instabilities leading to penetration of the
infalling plasma into pulsar magnetosphere, the pressure of the accreting gas
exceeds the magnetic pressure of the dipole with $B=10^{12}$ Gs. already after
one day of accretion at subcritical accretion rate $10^{-9}\,\,M_{\odot}$/year,
and the original magnetic dipole is completely
graved under a plasma layer, and currents in the plasma prevent external
appearance  of stellar magnetic field.
In reality the instability and gas penetration through the magnetosphere make the
screening process much slower, and when the penetration layer reaches the
surface of the star the external magnetic field strength of the neutron star
could reach the stationary state.

 From the statistical analysis of 24
binary radio pulsars with nearly circular orbits and low mass companions,
it was discovered \cite{vdh95} a clear correlation between spin period $P_p$ and
orbital period $P_{orb}$, as
well as between the magnetic field and orbital period: pulsar period
and magnetic field strength increases with the orbital period at
$P_{orb}>100$ days, and scatters around $P_p\sim 3$ ms and $B\sim 2\cdot 10^8$ Gs
for smaller binary periods. These relations strongly
suggest that an increase in the amount of accreted mass leads to a decay of the
magnetic field, and a 'bottom' field strength of about 10$^8$ G is also
implied. Several models of magnetic field screening during accretion had been
considered, see  \cite{cz98,cz00} and references therein.
Evolutionary aspects of formation of recycled pulsars are discussed in \cite{bisk74,ty73},
and summarized in \cite{bhav}.
Several scenario had been analyzed in \cite{bisk74,ty73} for the evolution of binaries,
massive enough to produce the neutron star or black hole in one or both remnants.
During the evolution such stars go through stages of different nuclear burning, which
are ended by formation  of carbon-oxygen white dwarf, or proceeds until formation of
the iron - peak elements core, which collapses with formation of the neutron star
or a black hole. The first born relativistic star (neutron, or black hole) goes through
the stage of an accreting binary X-ray source, and the evolution of such system is
finished after end of evolution of the second component, and the end of accretion.
During the evolution of massive close binary there are possibilities of a disruption
of the binary after the first or the second collapse, which may be accompanied by
explosion with large mass ejection, or by a kick due to anisotropy in a mass ejection
or in neutrino flux.
It was nevertheless concluded in \cite{bisk74}, that
"evolutionary analysis cannot exclude formation
of close binary, containing neutron star, and another star on the last evolutionary state
(white dwarf, neutron star, black hole). Contrary, this analysis indicates to formation
of such binaries at the end of evolution of sufficiently massive
systems with a large probability." Same conclusion was obtained in \cite{ty73}.
The absence of discovery of binary pulsars in 1973
was unambiguously explained in \cite{bisk74} as follows:
"The absence of radiopulsars in close binaries
may be explained if we suggest additionally, that magnetic field of the neutron star is
damped during accretion  stage"

The probabilistic conclusion about formation of radiopulsars in close binaries was
definitely confirmed in \cite{bisk74} by the analysis of the evolution of
one of the best learned X-ray pulsars Her X-1. The component of the neutron star
in this system is a low mass star which will end its life by formation of the
white dwarf, when the pair cannot be disrupted. On the other hand, during
accretion the neutron star acquires a rapid rotation, and it was shown in
\cite{bisk74}, that during the process of termination of accretion the neutron
star - X-ray pulsar cannot follow the stationary period, increasing with
decreasing of the the accretion rate $\dot M$, and after the end of accretion
will remain to be rather rapidly rotating neutron star with a period
of the order, or smaller than periods of the known radiopulsars. So, the only
explanation of absence of binary radiopulsars in to 1973 year remains their
faintness, connected with low magnetic field.

The first proof of this conclusion came soon by discovery of the first binary
radiopulsar PSR 1913+16 \cite{ht75}. This pulsar was a rapidly rotating with a period 0.059 s
in the close binary with a period 7$^h$45$^m$. Just after the discovery I
have identified this object with the old recycled pulsar with low magnetic field
\cite{bkp}. Measurements \cite{T76}
of the period derivative $\dot P$, which gave possibility
to estimate the age of this pulsar $\tau=\frac{P}{2\dot P}\approx 10^8$ years,
and magnetic field $B\approx 2 \cdot 10^{10}$ Gs had proofed this identification.
Among more than 100 recycled pulsars most objects consist of a neutron star with low mass
white dwarf companion, and the systems with two neutron stars, like in PSR 1913+16
are in minority.
Double-neutron-star (DNS) binaries are rare, and only six such systems
are known. All recycled pulsars, with white
dwarf or neutron star companions, have magnetic fields $10^8\, -\, 10^{10}$,
so the hypothesis about field damping on the accretion stage seems to be correct.

Another convincing evidence in favor of magnetic field graving during accretion
appeared by discovery of the first binary system, containing two
radiopulsars.
The recently discovered 23-ms pulsar J0737$-$3039A  was found to be in a 2.4-hr
eccentric orbit with another compact object that the observed orbital
parameters suggested was another neutron star \cite{bdp+03}.
 The 2.8-sec pulsar J0737$-$3039B was discovered \cite{lyn04}
 as the companion to the  pulsar J0737$-$3039A in a
highly-relativistic double-neutron-star system, allowing unprecedented
tests of fundamental gravitational physics.
The short orbital
period and compactness of the system and the high timing precision
made possible by the large flux density and narrow pulse features of
this pulsar promise to make this system a superb laboratory for the
investigation of relativistic astrophysics. The resulting in-spiral will end in
coalescence of the two stars in about 85~My.  This discovery
significantly increases the estimates of the detection rate of DNS
in-spirals by gravitational wave detectors \cite{bdp+03}.
The clock-like properties of pulsars moving in the gravitational
fields of their unseen neutron-star companions have allowed unique
tests of general relativity and provided evidence for gravitational
radiation.

The properties of these pulsars fit very well the evolutionary scheme
of formation of binary pulsars from \cite{bisk74}. According to this scheme the
23 ms pulsar had accelerated its rotation during accretion, when its magnetic field
was decreased due to screening of the accreting material. The second 2.8 s pulsar
was born later in SN explosion, which did not disrupt the pair. Therefore, ms pulsar
should be older, and have weaker magnetic field. Indeed, 23 ms pulsar has
$B= 6.3\times 10^9$ Gs, and characteristic age $\tau= 210$ My; and 2.8 s pulsar has
 $B=1.6\times 10^{12}$ Gs, and $\tau= 50$ My \cite{lyn04}.
The masses of neutron stars in this binary are estimated as
1.34 M$_\odot$ for 23 ms, and 1.25 M$_\odot$
for 2.8 s pulsars.

\subsection{Magnetic field damping during accretion}

Accretion induced Ohmic heating and dissipation  have been investigated in many works
(see i.g. references in \cite{chou02}).
This mechanism may work only if the magnetic field is produced by electrical current
in the outer layers of the accreting neutron star, which may be heated during accretion.
Evidently, the highly degenerate neutron star core is not sensible to this heating
and will preserve its very high conductivity. Therefore magnetic field, originated by deep
inside electrical currents may become weaker only due to screening by the infalling plasma
during accretion \cite{bisk74}.
Different models of magnetic field screening during accretion have
been considered. It is usually accepted that matter is channelized
by the dipole magnetic field and flows to the magnetic poles of
the neutron star \cite{bisf69,ag}.
  As the magnetic
field of the neutron star decreases because of the screening due to
accreting material, it is less able to channelize the accretion flow
and thereby the polar cap widens.  One can easily find out how the
angular width $\theta_{\rm P}$ of the polar cap depends on the surface
magnetic field $B_{\rm s}$ of the neutron star (see, for example, \cite{shap}).
The field line starting from $\theta_{\rm P}$
at the surface of the neutron star, with a radius $r_{\rm s}$, is the last
closed field line of the dipolar field and passes through the Alfv\'en
radius $r_{\rm A}$. It easily follows that

\begin{equation}
\label{ch1}
\sin \theta_{\rm P} = \left(\frac{r_{\rm s}}{r_{\rm A}} \right)^{1/2}.
\end{equation}
Assuming that the ram pressure of the freely in-falling accreting material
at the Alfv\'en radius equals the magnetic pressure, a few steps of
easy algebra give

\begin{equation}
\label{ch2}
r_{\rm A} = (2GM)^{-1/7} r_{\rm s}^{12/7} B_{\rm s}^{4/7} {\dot M}^{-2/7},
\end{equation}
where $M$ is the mass of the neutron star and $\dot M$ the accretion rate.
It follows from (\ref{ch1}) and (\ref{ch2}) that

\begin{equation}
\label{ch3}
\sin \theta_{\rm P} \propto B_{\rm s}^{-2/7}.
\end{equation}
This is how the polar cap widens with the weakening magnetic field until
$\theta_{\rm P}$ becomes equal to $90^o$ when (\ref{ch3}) obviously ceases to hold.
On taking $M = 10^{33}$ gm, $\dot M = 10^{-8} M_{\odot}$ yr$^{-1}$,
$r_{\rm s} = 10$ km, $B_{\rm s} = 10^{12}$ G, we find from (\ref{ch2}) that
$r_{\rm A} \approx 300$ km.  Substituting this in (\ref{ch1}), we conclude that
the initial polar cap angle is of order $10^o$.

The accreting materials falling through the two polar caps flow horizontally
towards the equator in both the hemispheres.  At the equator, the
opposing materials flowing in from the two poles meet, sink underneath the
surface (inducing a counter-flow underneath the equator-ward flow at
the surface) and eventually settle radially on the neutron star core.
With a suitably specified flow having these characteristics, it was studied
 kinematically in
\cite{kon04,chou04} how the magnetic field evolves with time, taking
into account the fact that the polar cap width changes with the evolution
of the magnetic field, thereby altering the velocity field also.
It was found in the simulations \cite{chou04}
that the equator-ward flow near the surface is quite
efficient in burying the magnetic field underneath the surface. However, when
the polar cap opens to $90^o$, the accretion becomes spherical and radial.
It is found that such accretion is not efficient in burying
the magnetic field any further. The magnetic field at the surface of the neutron
star keeps decreasing until the polar cap opens to $90^o$, after which the
magnetic field is essentially frozen, since the radial accretion cannot
screen it any further. If $\theta_{\rm P,i}$ is the initial polar cap
width, then it follows from (\ref{ch3}) that the magnetic field would decrease
by a factor $(\sin 90^o/ \sin \theta_{\rm P,i})^{7/2}$ from its initial
value before it is frozen to an asymptotic value.  On taking $\theta_{\rm P,i}$
in the range $5^o$--$10^o$, this factor turns out to be about $10^3$--$10^4$,
exactly the factor by which the magnetic fields of millisecond pulsars
are weaker compared to the magnetic fields of ordinary pulsars.

Put another way, the magnetic field freezes when the Alfv\'en radius
becomes equal to the neutron star radius.  The asymptotic value of the
surface magnetic field can be found directly from (\ref{ch2})
by setting $r_{\rm A}$
equal to $r_{\rm s}$, which gives

\begin{equation}
\label{ch4}
B_{\rm asymp} = (2 G M)^{1/4} \dot M^{1/2} r_{\rm s}^{-5/4}.
\end{equation}
On using the various standard values mentioned before, we
find $B_{\rm asymp} \approx 10^8$~G. When the magnetic field falls to
this value, it can no longer channelize the accretion flow, resulting in
the flow becoming isotropic.  Such a flow is unable to screen the
magnetic field any further.  After the accretion phase is over, the neutron
star appears as a millisecond pulsar with this magnetic field.
This could be the reason why
millisecond pulsars are found with magnetic fields of order $10^8$ G.
Note that the accretion flow in the polar region of a magnetized
neutron star is sensitive to various magneto-hydrodynamic
instabilities and it is difficult to make an assessment
of the effectiveness of screening without a full three-dimensional
computation which is yet to be attempted.

A simple analytical model of the magnetic field screening in incompressible
fluid approximation was considered in \cite{cz98,cz00}. It was found a very rapid
decrease of the magnetic field in the polar cap during a time scale \\
$\tau_p \approx 10^5 {\dot M} (m_B/10^{-3}M_{\odot})$ years, where
$m_B=(B_{\rm asymp}/B_0)^{4/7}M_{cr}$ with $B_{\rm asymp}$ from (\ref{ch4}),
initial magnetic field of the neutron star $B_0$, and mass of the crust
$M_{cr}=0.03 M_{odot}$. Here the polar magnetic field is tending to the
same asymptotic value (\ref{ch4}), and development of instabilities should
decrease also the equatorial magnetic field to similar values.

It was noted in \cite{bisk76}, that "magnetic fields buried during intensive
accretion may start to "percolate" outside after the end of accretion", what
could increase the magnetic field of recycled pulsars with time, see also
\cite{mic94}. Some estimations of the field percolation outside have been done
in \cite{mus95}. The authors considered cases with very rapid field increase
due to percolation during
$10^3 -- 10^4$ years, up to the initial value which the neutron star had before
the start of accretion. Evidently, this conclusion is in contradiction with observation,
because all recycled pulsars which could be very old, even up to $10^{10}$ years
\cite{lor03}, have equally small magnetic fields. This contradiction is connected with
the artificial model considered in \cite{mus95}, where the percolation was started
from the depth not more than 260 m, corresponding to the accreted material
less that $5 \cdot 10^{-5} M_{\odot}$. The conductivity in the outer crust of the neutron star
could be small enough to permit such a rapid percolation.
Actually, during the accretion phase lasting $10^7 -- 10^8$ years the amount of the accreted
material is much larger, the magnetic field is buried to much deeper layers with
much higher conductivity. Therefore the outward percolation of the magnetic field in the
recycled pulsars should be negligible.

\subsection{Formation of single recycled pulsars: Enhanced Evaporation}

Among the pulsars, which can be attributed to the recycled (rapid rotation + low magnetic field)
there are many single objects.
The mechanism of the pair disruption and formation of a single RP is not quite clear.
After discovery of the first eclipsed radiopulsar it was suggested that the companion
could be evaporated \cite{klu88}, but statistical properties did not confirm it.
The concentration of the recycled pulsars
(RP) to the globular clusters (GC), 47 RP from the total number 103 are situated in
GC \cite{lor03}, is very similar to the distribution of low mass X-ray binaries (LMXB).
Taking into account, that presently the total relative mass of GC in Galaxy is about
$10^{-3}$ and about one half of LMXB are situated in GC, the relative concentration of
LBXB in GC is 1000 larger than in the Galaxy \cite{sh82}. The
similarity of distributions of LMXB and RP may be considered as another evidence
to their genetic relation \cite{bisk74,bisk76,rs83}. According to \cite{lor03} there are
21 single RP in GC (45\%), and only 9 single RP in the galactic bulge (16\%). This evident
statistical difference indicates that not only birth of LMXB, but also the disruption
of the binaries are connected with close stellar encounters, which are much more frequent
in GC than in the bulge \cite{bk89}. It was suggested in \cite{bk89}, that single RP
in the bulge could be the remnants of the completely evaporated GC, same origin may have
also LMXB of the bulge \cite{bkr83}.
Numerical simulations made in \cite{rap89} have shown that only
pairs with large orbital periods 10 -- 100 days could be destroyed by close stellar encounters
in globular clusters. However, it follows from observations \cite{lor03}, that RP are usually
members of more close binaries. The solution of this problem may be done by the mechanism of
"Enhanced Evaporation" (EE), suggested in \cite{bk90}.

It was shown theoretically in \cite{gl50} and confirmed later by numerical experiments
\cite{aa75}, that stellar encounters of stars with close binaries lead to energy extraction
from the binary and heating of the cluster field stars. Contrary to that, collisions with
sufficiently wide binary lead to its father widening, and finally to its destruction.
This property is interpreted very easily. The kinetic energy exchange between stars
during encounter is determined by their velocities, almost independently on their
single or binary origin. Relaxation processes in stellar encounters lead to establishing of
equipartition of energies of all stars. It means, that star with larger kinetic energy lose it,
and the one with smaller energy gain it. For single stars it would lead to real equipartition,
so that stars with larger energies in average loose it, but contrary result takes place if one
of the encountering objects is a close binary. In close binary star is rapidly rotating on the
orbit. If this rotational energy exceeds the average energy of stars in the cluster there is
an average energy transfer to the field stars. When the star in the binary loose its
kinetic energy, its orbit is changing in such a way that is starts to rotate faster than before,
and the pair becomes tighter. The binary has an effective negative heat capacity, connected
with the validity of the virial theorem \cite{ll62} in the binary system. The net loose of
the energy by the pair leads to acceleration of the orbital motion. When tight binaries are
formed in the GC, they could destroy it during time less than the cosmological one, so the
LMXB and single RP in the bulge could be formed in such a way.

The situation is quite different, when one of the degenerate members in the close binary
fills its Roche Lobe. Let isolated binary contain a neutron star with a mass $M_1$,
and a degenerate dwarf with a mass $M_d$. When the dwarf fills its Roche lobe, the radius
of the dwarf $R_d$ is connected with a distance between stellar mass centers
$R_{12}$ by a relation

\begin{equation}
\label{jup1}
R_d=R_{12}\frac{2}{3^{4/3}} \left(\frac{M_d}{M}\right)^{1/3}, \quad
M=M_1+M_d.
\end{equation}
For white dwarfs with an equation of state valid in small mass dwarfs the
equilibrium models had been constructed in \cite{zs69}.
The radius of a degenerate dwarf increases with decreasing of its mass until
very low masses, when repulsion effects may change this dependence.
Gravitation radiation leads to
loss of the angular momentum of the system, and tries to make the binary closer, but
approach of the dwarf to the neutron star induces a mass transfer. If we suggest that the
dwarf fills its Roche lobe, which radius coincides with the dwarf radius, than due to
mass transfer dwarf mass decreases, its radius increases with the Roche lobe radius, and
the binary is becoming softer. Evolution of such binary under the action of
a gravitational radiation was calculated in \cite{bk90}. It was obtained that during
a cosmological time $\tau_c=2\cdot 10^{10}$ years the mass of the carbon dwarf
reaches $0.0025M_{\odot}$. At that time the period of the binary $P$ reaches about $1.5$ hours,

\begin{equation}
\label{jup2}
P=\frac{2\pi R_{12}^{3/2}}{(GM)^{1/2}}=
\frac{9\pi R_d^{3/2}}{\sqrt{2G M_d}},
\end{equation}
and the dwarf velocity in the binary
\begin{equation}
\label{jup3}
v_d=\frac{2\pi R_{12}}{P}=\frac{\sqrt{2}}{3^{2/3}}
\left(\frac{GM^{2/3}M_d^{1/3}}{R_d}\right)^{1/2}, \quad
{\rm at}\,\,\, M_d \ll M.
\end{equation}
The pair becomes soft, and would be destroyed by close encounters with the field
stars when the kinetic energy of the dwarf in the binary is less than the average
kinetic energy of the field stars: $M_f v_f^2 > M_dv_d^2$. For globular clusters
47 Tuc and $\omega$Cen the corresponding velocities are 10.3 and 16.8 km/s, and
masses are equal to $0.67 M_{\odot}$ and $0.51 M_{\odot}$ respectively.
According to calculations \cite{bk90}, the pair in these GC
is becoming soft, when the dwarf mass becomes less than $6 \cdot 10^{-4}$ and
$9 \cdot 10^{-4}$ respectively. These masses are several times less than the dwarf mass
after cosmological time, and to reach these masses one needs about 10 cosmological times.

It was shown in \cite{bk90}, that stellar encounters
of the field stars with "hard" pairs,
when the binary loose its energy and momentum,
act similar to the gravitational radiation. They make the hard pair even harder,
but when the dwarf in the pair fills its Roche lobe such encounters make the pair
softer. When the encounters with the field stars are effective enough they may transform
the hard binary in the state of the mass exchange into the soft one, which would be destroyed
directly by close encounters. This process was called in \cite{bk90} as
"enhanced evaporation".
For the estimation of the momentum relaxation time $\tau_m$ we use the expression \cite{spi},
which is valid for stars with comparable masses, because in the close binary
the centrum mass motion may take the extra momentum

\begin{equation}
\label{jup4}
\tau_m=\frac{v_d^3}{4\pi G^2 M_f^2 n_f \Lambda}.
\end{equation}
Here Coulomb logarithm $\Lambda \approx 10$, $n_f$ is the average density of
the field stars. With account of (\ref{jup3}) the characteristic encounter time,
which determines the time of the pair destruction due to EE is written as

\begin{equation}
\label{jup5}
\tau_m\approx 10^{22}\frac{m m_d}{m_f^2 n_{51}}\,\,\, {\rm c} .
\end{equation}
Here small $m$ determine masses in solar units, $n_{51}=n_f/10^{-51}$ cm$^{-3}$.
The approximate relation for dwarf mass - radius relation is used, which is valid for
the ideal fermi gas stars, $R=7.6\cdot 10^8 m_d^{1/3}$. Everywhere carbon dwarfs with
$\mu_e=2$ are considered, $\mu_e$ is the number of baryons per one electron.
Following the evolution of the pair due to encounters it was found in \cite{bk90} that
in order to destroy the binary due to EE process the field star density should exceed

\begin{equation}
\label{jup6}
n > 3\cdot 10^5 \frac{m^{9/11}}{m_f^2} {\rm pc}^{-3}.
\end{equation}
The central parts of the most dense GC, where single recycled
pulsars are concentrated, like M15, satisfy this demand.
So, EE process is probably responsible for the formation of the single RP in GC, as
well as in the bulge, where they could  find themselves after the evaporation
of the whole GC. Note, that binaries with the hydrogen-helium brown dwarfs are
destroyed easier due to EE \cite{bk90}, with the coefficient $2$ instead of $3$ in
(\ref{jup6}).

\section{X-ray binaries}

There are several ways to estimate observationally magnetic
field of an X-ray pulsar. During accretion matter is stopped by the
magnetic field at the alfvenic surface, where gaseous and magnetic pressures
are in a balance. At stationary state Keplerian angular velocity of
the accretion disk at the alfvenic surface is equal to the stellar
angular velocity \cite{pr72} $\Omega_K=\Omega_s=\Omega_A$.
Otherwise neutron star would be accelerated due to
absorption of matter with large angular momentum, or decelerated due
to throwing away matter with additional angular momentum \cite{is75}.
X-ray pulsars may have spin-up and spin-down
stages \cite{bil}, but most of them show average
spin-up, what indicate to
their angular velocity being less than the equilibrium one.
This is observed in the best studied X-ray
pulsar Hex X-1 \cite{ska92,det98}. Analysis of spin-up/down phenomena in the
X-ray pulsars indicate to important role of the mass loss \cite{lrbk95},
and to stochastic origin of spin-up/down transitions \cite{lrbk99}.

For a given luminosity $L_{36}=L/(10^{36}$ergs/s) and dipole magnetic
field at stellar equator $B_{12}=B/(10^{12}$Gs) we get the following
value of the equilibrium period \cite{lip92}, for the neutron
star mass $M_s=1.4 M_{\odot}$

\begin{equation}
\label{ref3}
P_{eq}\approx 2.6\,B_{12}^{6/7}\,L_{36}^{3/7} {\rm s}.
\end{equation}
For Her X-1 parameters $L_{36}=10,\,\,P=1.24$ s, we get a magnetic
field corresponding to the equilibrium rotation $B_{12}^{eq}=0.9$.
Taking into account the average spin-up of the pulsar in Her X-1, we
may consider this value as an upper limit of its magnetic field.
Even more rough estimations of the magnetic field in X-ray pulsars
follow from the average spin-up rate under condition $\dot J_{rot}=
\dot M\,\Omega_A$, or restrictions on the polar magnetic field value
following from the observed beam  structure and condition of local
luminosity not exceeding the critical Eddington one. These conditions
lead to smaller values of the magnetic field of Her X-1 on the level
$10^9\, -\, 10^{10}$ Gs \cite{bisk74,bisk75,bis74}.

Observations of low mass X-ray binaries (LMXB)
indicate to very low values of their magnetic fields due to absence
of X-ray pulsar phenomena. Modulation of X-ray flux permitted to
reveal the rotational period of the neutron star in the LMXB
SAX J1808.4-3658 corresponding to the frequency 401 Hz, due to
RXTE observations (see reviews \cite{vdk98,w04}). This observations
fill a gap and form a long-waiting link between LBXB and recycled
millisecond pulsars \cite{rs83}, as neutron stars with very low
magnetic field (up to $10^8$ Gs).

The most reliable estimation of the magnetic field of the X-pulsar Her X-1
comes from detailed observations of the beam variation in this pulsar on
different stages, made on the satellites ASTRON \cite{ska92},
and GINGA \cite{det98} and RXTE \cite{sco20}. This pulsar, in addition to 1.24 s
period of the neutron star rotation, is in a binary system with an orbital
period 1.7 days, and shows a 35 day cycle, where during only 12 days its
luminosity is high. During other 23 days its X-ray luminosity strongly
decreases, but small changes in the optical luminosity and remaining
strong reflection effect indicate, that the X-ray luminosity
remains almost the same during all 35 day cycle. Visible
decrease of the X-ray flux is due to an occultation phenomena.
The model which explains satisfactory the phenomenon of the 35 day cycle is
based on the precession of the accretion disc with the 35 day period, and
occultation of the X-ray beam during 23 days. Analysis of the beam
structure during high and low X-ray states lead to the conclusion, that
during the low state we observe not the direct X-ray flux from the
neutron star, but the flux, reflected from the inner edge of the
accretion disc. This conclusion is based on the $180\deg$ phase shift
between the X-ray beams in high and low states \cite{ska92,det98,sco20}.
In order to observe the X-ray flux reflected from
the inner edge, situated near the Alfven radius of the accretion disc,
it cannot be very far away from the neutron star. The
estimations give the upper limit to the ratio of the Alfven and stellar
radiuses

\begin{equation}
\label{ref4}
\frac{r_A}{r_s} \leq 20.
\end{equation}
The schematic picture of the accretion disc and its inner edge orientation
around the neutron star at different stages of the 35 day cycle
is shown in Fig.1, taken from \cite{ska92}. As was indicated
above, the value of the Alfven radius is determined by the neutron star mass
$(M_s=1.4 M_s)$, mass flux $\dot M=3 \times 10^16$ g/s, corresponding to
the luminosity $L=10^{37}$ ergs/s, and the value of the magnetic field.
Taking dipole radial dependence of the magnetic field $B=B_s(r_s/r)^3$,
and neutron star radius $r_s=10^6$ cm, we obtain the ratio
in the form $r_A/r_s\approx 300 B_{12}^{4/7}$. To have this ratio not
exceeding 20 we get an inequality $B \leq 3\times 10^{10}$ Gs.

\begin{figure}
\centerline{\hbox{\includegraphics[width=0.8\textwidth]{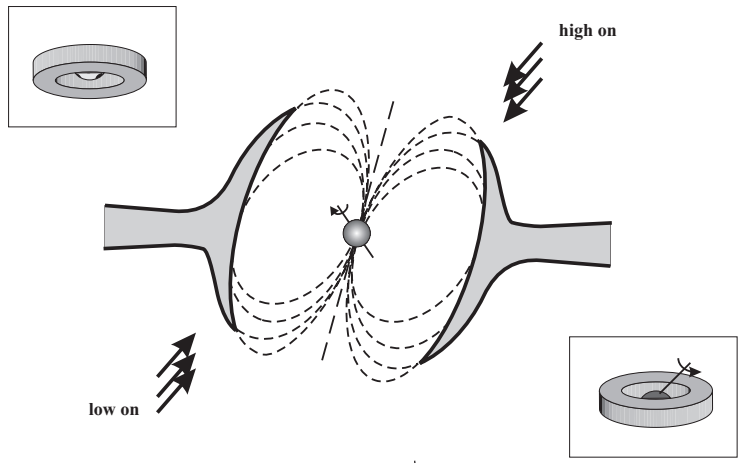}}}
\caption[h]{Configuration of the inner edge of the disk and the neutron star;
neutron star and the disk in the "high-on" state (left top box),
and in the "low-on" state (right bottom box).}
\end{figure}

It was found in \cite{tp78} a feature in the spectrum of Her X-1
at energies between 50 and 60 keV. Interpreting it as a cyclotron line
feature according to $E_X=\frac{\hbar e B}{m_e c}$ leads to the
value of the magnetic field
$B_{cycl}=(5-6)\times 10^{12}$ Gs, what is much higher
than any other above mentioned estimations. Spectral features had been
observed also in other X-ray sources. Recent observations on RXTE
\cite{he99}, and Beppo-SAX \cite{b}
of the pulsating transient 4U 0115+63 had shown a presence of
3 and 4 cyclotron harmonics features, corresponding to the
magnetic field strength of $1.3\times 10^{12}$ Gs.
A comparison of the shapes of the beam in cyclotron harmonics may
be used for testing the nature of these features.
Cyclotron features had been observed in several X-ray sources
\cite{mmn97}, and
they always had corresponded to large values of magnetic fields
$B_{cycl}>10^{12}$ Gs. Such situation was not in good accordance with
a well established observational fact, that all recycled pulsars,
going through a stage of an X-ray source, have much smaller magnetic
fields, usually not exceeding $10^{10}$ Gs. Particularly, for the Her X-1
the value of its magnetic field, following from the cyclotron interpretation
of the spectral feature, was in contradiction with all other observations,
including the most reliable, based on the beam shape variability
during 35 day cycle.

\section{Relativistic dipole interpretation of
        the spectral feature in Her X-1}

It seems likely that this conflict is created by using the
non-relativistic formula connecting cyclotron frequency with a value of
the magnetic field. According to \cite{bisf69},
ultrarelativistic electrons with a temperature $\sim 10^{11}K$,
may be formed in the non-collisional shock during accretion,
emitting a relativistic dipole line.
The mean energy of this line is broadened and shifted
relativistically, in comparison with the cyclotron line,
by a factor of $\gamma \simeq \frac{kT}{mc^2}$.
The spectrum profile of the relativistic dipole line is calculated
in\cite{babk} for
various electron distributions, where the model of the hot spot
 of Her X-1 is considered, and it is shown that the overall observed
X-ray spectrum (from 0.2 to 120 KeV) can arise under the fields near
$5\cdot 10^{10}$~Gs which are well below $B_{cycl}$, and are not
in the conflict with other observations.

According to \cite{gs73,bis73},
in the magnetic field near the pulsar the cross
component of a momentum of electrons is emitting rapidly,
while the parallel velocity remains constant.
Hence the momentum distribution of the electrons is anisotropic
 $p_\perp^2\ll p_\parallel^2$, with
$p_\perp\ll mc$, $p_\parallel\gg mc$.
Assume for simplicity that the transverse electron
distribution is two-dimensional Maxwellian
$
  dn=\frac{N}{T_1}{\exp \left( -\frac{m u^{2}}{2T_1} \right)}\,
  d\frac{m u^{2}}{2}
$, $T_1 \ll mc^2$.

In the relativistic dipole regime of
radiation the electron is non-relativistic in
the coordinate system, connected with the Larmor circle. In the
laboratory system, where the electron is moving with a velocity $V$,
the angle between the magnetic field and electron momentum
vectors is smaller than the angle of the emitting beam. In this conditions
we may consider that all radiation is emitted along the magnetic field with
an intensity $J(0)$, after integration over $du$,
at frequency $\omega_{md}$  as \cite{babk}

\begin{equation}
\label{ref5}
J(0)=\frac{2 N e^4 B^2 T_1}{\pi c^5 m^3 (1-\frac{V}{c})},\qquad
\omega_{md}=\omega_{cycl} \sqrt \frac{1+\frac{V}{c}}{1-\frac{V}{c}}
\approx 2 \omega_{cycl}\frac{E_\parallel}{m_e c^2},
\end{equation}
where
$\omega_{cycl}=\frac{eB}{m c}$.
That gives
$1-\frac{V}{c}=\frac{2 \omega_{cycl}^2}{\omega^2}$.
Let us consider the parallel momentum distribution of the electrons as:
$dn=f(p_\parallel)\,dp_\parallel$.
Substituting of $dn$ for $N$ and using
$p_\parallel=\frac{mc}{2}\frac{\omega}{\omega_{cycl}}\,$,
we obtain for the spectral density

\begin{equation}
\label{ref6}
J_\omega=\frac{e^2 T_1}{2 \pi c^2 \omega_{cycl}} \omega^2
 f \left( \frac{m c }{2}\frac{\omega}{\omega_{cycl}}\right )\,d\omega.
\end{equation}
Let us consider two cases. The first is a relativistic Maxwell
$f=\frac{n_0 c}{T_2} \exp\left( -\frac{p_\parallel c}{T_2}\right )$,
where $n_0$ is a number of emitting electrons. The spectrum is

\begin{equation}
\label{ref7}
J_\omega=\frac{n_0 e^2}{2 \pi c \omega_{cycl}}\frac{T_1}{T_2} \omega^2
 \exp\left( -\frac{m c^2 \omega}{2 \omega_{cycl} T_2}\right )
\,d\omega.
\end{equation}
It has a single maximum at
$\frac{\omega}{\omega_{cycl}}=\frac{4 T_2}{m c^2}$.
In the second case we have
$f=\frac{n_0 }{\sqrt {\pi} \sigma}
\exp\left[ -\frac{{(p_\parallel-a)}^2}{\sigma^2}\right ]$.
The spectrum of radiation is

\begin{equation}
\label{ref8}
J_\omega=\frac{n_0 e^2 T_1}{2 \pi^{3/2} c^2 \omega_{cycl}\sqrt{\sigma}}
\omega^2
 \exp\left( -\frac{(\frac{m c}{2}\frac{\omega}{\omega_{cycl} }-a)^2}
 {\sigma^2}\right )
\,d\omega.
\end{equation}
When $\sigma \ll a$ this spectrum has a single maximum at
$\omega \simeq \frac{2 a}{m c} \omega_{cycl}$.
\cite{babk}
had approximated experimental spectra taken from \cite{mmo90,mcb82}.
The last spectrum (solid line)
and its fitting (dashed line) are shown in Fig.2.
It was taken in accordance with \cite{bisf69},
$a=7 \cdot 10^{-4}$~${\rm\frac{eV \cdot s }{cm}}$, corresponding to
average electron energy $E_{\parallel}=ac \approx 20$ MeV,
and the best fit for the line shape was obtained at
the magnetic field strength $B=4\cdot 10^{10}$~Gs.
In this model the beam of the "cyclotron" feature is determined by the
number distribution of the emitting relativistic electrons,
moving predominantly
along the magnetic field, over the polar cap.

\begin{figure}
\centerline{\hbox{\includegraphics[width=0.8\textwidth]{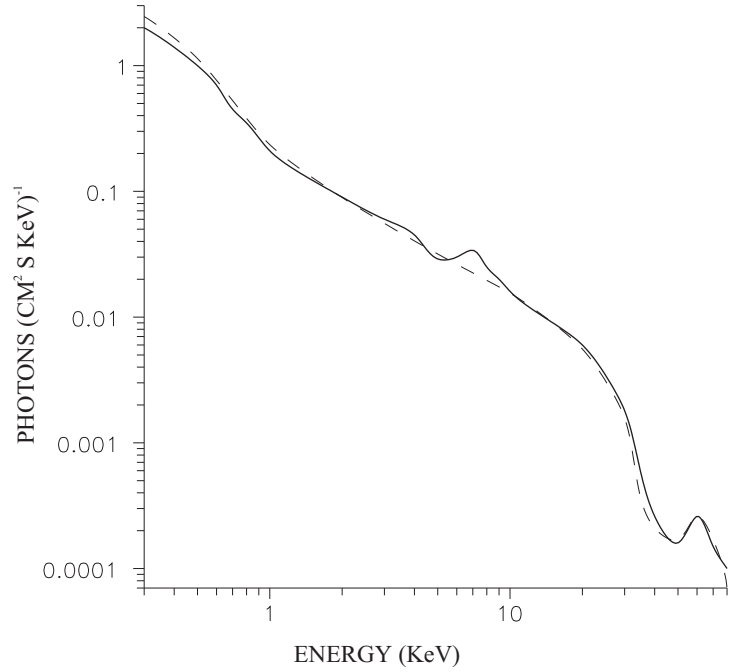}}}
\caption[h]{
  Comparison of the observational and computational X-ray spectra
of Her X-1. The solid curve is the observational results taken from
\cite{mcb82},
the dot curve is the approximation with
$T_s=0.9$~KeV, $T_e=8$~KeV, $\tau_e=14$,
$a=7 \cdot 10^{-4}$~${\rm\frac{eV \cdot s }{cm}} $,
$\sigma=10^{-4}$~${\rm\frac{eV \cdot s }{cm}} $, $B=4 \times 10^{10}$ Gs.
}
\end{figure}

In order to obtain the whole experimental spectrum of the Her X-1
the following model of the hot spot (Fig.3) was considered
in \cite{babk}.
A collisionless shock wave is generated in the accretion flow
near the surface on the magnetic pole of a neutron star. In it`s front the
ultrarelativistic electrons are generated.
Under the shock
there is a hot turbulent zone with a temperature $T_e$,
and optical depth $\tau_e$, situated over a heated spot
with a smaller temperature on the surface
of the neutron star.

{\begin{figure}
\centerline{\hbox{\includegraphics[width=0.8\textwidth]{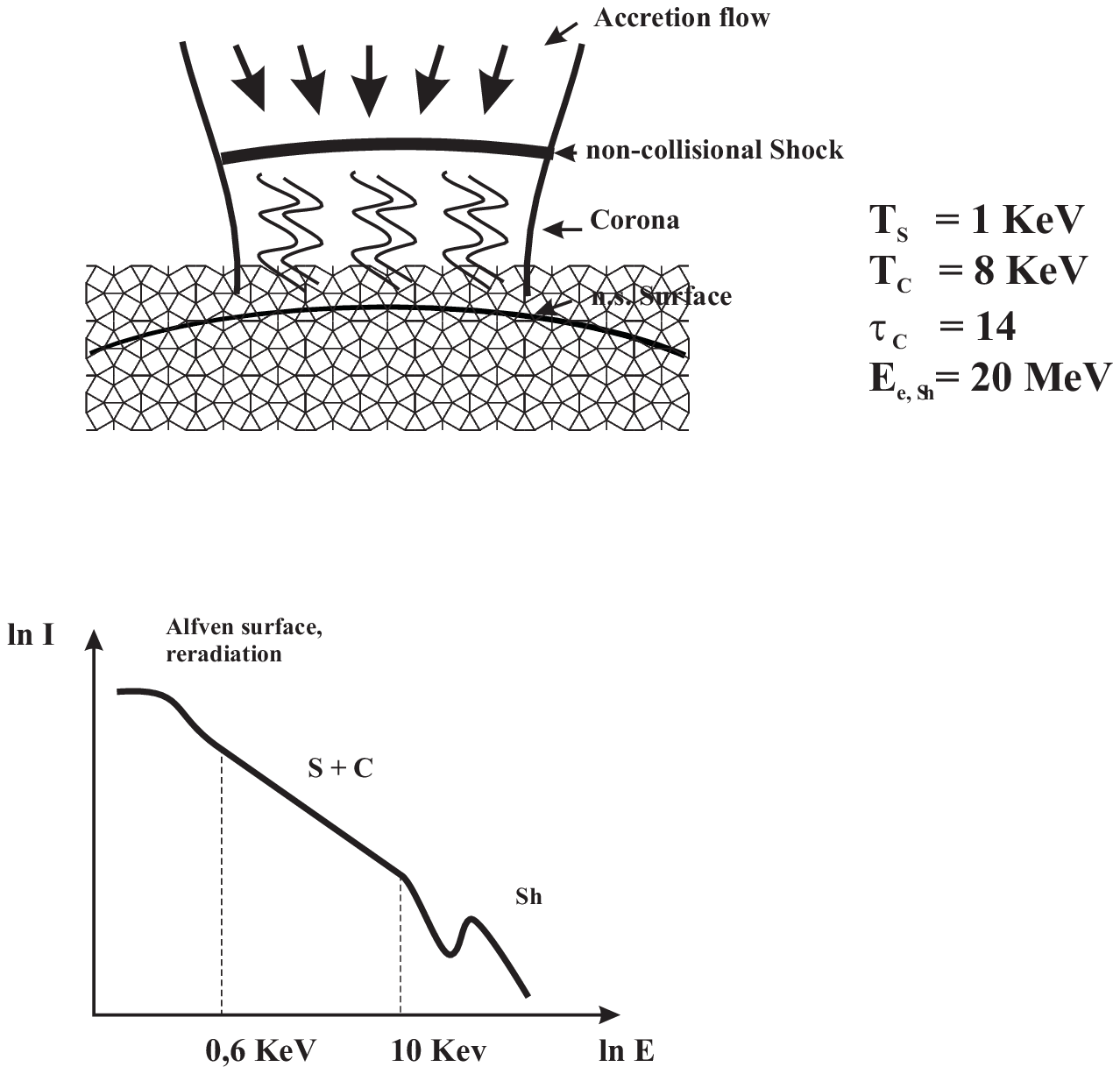}}}
\caption[h]{ Schematic structure of the accretion column near the
magnetic pole of the neutron star (top), and its radiation spectrum (bottom).
}
\end{figure}
}

The whole X-ray spectrum of pulsar Her X-1 from \cite{mcb82}
is represented in Fig.3 by the solid line.
There are three main regions in it:
a quasi-Planckian spectrum between 0,3 and 0,6 KeV, that is generated
(re-radiated) near the magnetosphere of the X-ray pulsar;
power-law spectrum $(0.6\div
20)$~KeV with a rapid decrease at 20 KeV, and the "cyclotron" feature.
The power-law spectrum is a result of comptonization in the corona
of a black-body spectrum emitted by the stellar surface.
The comptonized spectrum has been calculated according to
\cite{st80}.
Setting the neutron star radius equal to 10 km, distance to the
X-ray pulsar 6 Kps, hot spot area $S=2 \cdot 10^{12}$~cm$^2$,
the best fit was found
at $T_s=1$~KeV, $T_e=8$~KeV, $\tau_e=14$, which is
represented in Fig.3 by the dashed line.

The observations of the variability of the "cyclotron" lines
are reported by \cite{mmn97}. Ginga detected the
changes of the cyclotron energies
in 4 pulsars. The change is as much as 40 \% in the case of 4U 0115+63.
Larger luminosity of the source corresponds to smaller average energy of the
cyclotron feature. These changes might be easily explained in our model.
The velocity of the accretion flow decreases with increasing of the
pulsar
luminosity because locally the luminosity is close to the Eddington limit.
As a result the shock wave intensity drops as well as the energy of the
ultrarelativistic electrons in it`s front, leading to
decrease of the relativistic dipole line energy.

\section{Cyclotron harmonics in the X-ray pulsars}

Reports about discoveries of several cyclotron harmonics in spectra
of several X-ray pulsars appeared according to observations from RXTE and BeppoSAX:
  X0115+63 \cite{a}; Vela X-1 \cite{c,c1}; 4U1907+09
\cite{c2,c3}; A0535+26 \cite{d}
There are at least 3 harmonics observed in the spectrum of
 X0115+63 \cite{d1,b}.
This is the most reliable source with the cyclotron harmonics.
Other sources are less reliable, the second harmonic there is very weak.
These observations are very important and could help in solving
some problems of the theory of spectra formation in the X-ray pulsars.
There is still an open question, are the cyclotron features in the spectra
of X-ray pulsars of absorption or emission origin.
In the paper \cite{ba} all lines in 4 above mentioned sources are
interpreted as emission features.

It is not clear what kind of momentum distribution have electrons
radiating the cyclotron lines.
It is needed, that the motion of electrons
across the magnetic field would be subrelativistic (or nonrelativistic),
because otherwise many harmonics would merge, forming continuum synchrotron
spectrum. I most cases only one first harmonic is visible.
It is accepted usually that the speed of electrons along the magnetic field
is considerably less than the light speed $c$. However, it is possible,
like in the case of Her X-1, that the momentum distribution of the electrons is
highly anisotropic, so that their speed is subrelativistic across the magnetic field,
and ultrarelativistic along it.
Analysis of different observational data of the X ray pulsar Her X-1
had shown that all data can be explained without contradictions under
seggestion that the electrons forming the cyclotron feature are ultrarelativistic
with very anisotropic momentum distribution.

\begin{figure}
\label{f6}
\centerline{\hbox{\includegraphics[width=0.8\textwidth]{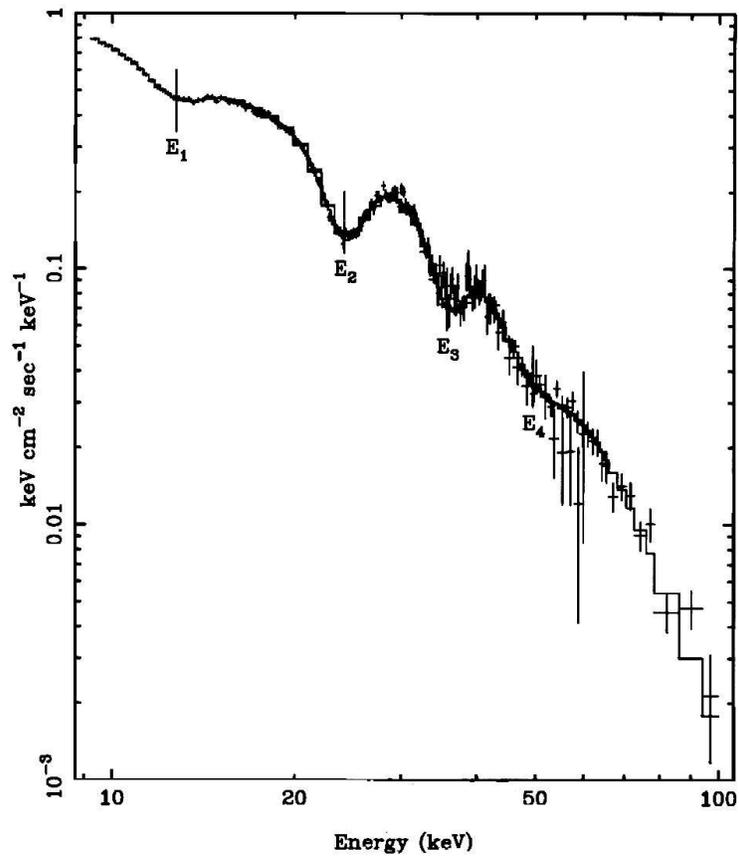}}}
\caption[h]{ Observational spectrum of the main pulse of
the X-ray pulsar X0115+63 in hard
part of the spectrum with harmonics, according to observation by BeppoSAX,
from \cite{b}}
\end{figure}

\begin{figure}
\centerline{\hbox{\includegraphics[angle=-90,width=0.8\textwidth]{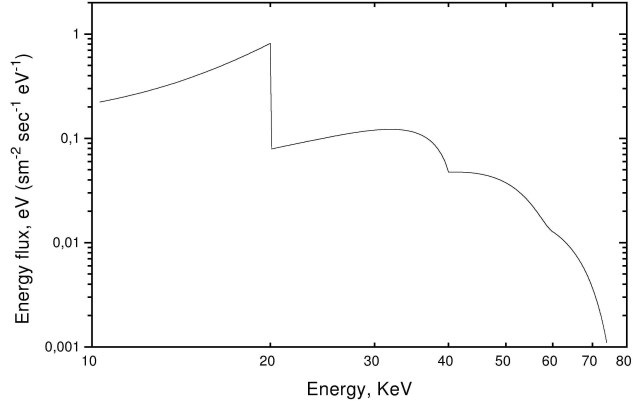}}}
\caption{Theoretical spectrum of 4 harmonics of magneto-dipole radiation
calculated according (\ref{f4}), for the perpendicular temperature
$T_{\perp}=20$~{ÊýÂ}. The non-dimensional energy is given along the $x$- axis in
a linear scale, and the emitted energy along the $y$ axis is given in the
logarithmic scale, from \cite{ba}.}
\label{f7}
\end{figure}

When the perpendicular electron energy is high enough for emission of several
cyclotron harmonics, the resulting spectrum of magneto-dipole radiation,
produced by electrons with nonrelativistic perpendicular and ultrarelativistic
parallel momentum distribution, has a very specific form \cite{ba}. Note that
the perpendicular energy is measured in the frame of the Larmor circle, and the
parallel energy is measured in the laboratory frame of the observer.
Different harmonics in the spectrum of magneto-dipole radiation become
nonequidistant in the laboratory frame. It could permit in principle to
determine are there electrons with the ultrarelativistic parallel energy
by analyzing the shape of the lines.
The number of photons of $n$-th harmonic, averaged aver the time,
which the emitted by the star in the unit of time, in the unit of the body angle,
in the unit frequency interval is determined  by the formula \cite{ba}

\begin{equation}
\label{f4}
 Q_n= \alpha \rho R^2 \frac{n^{2n}}{n!}
 \left(2\frac{\tilde \omega}{n}-
 \left(\frac{\tilde \omega}{n}\right)^2\right)^{n-1}
 \left(1+\left(\frac{\tilde \omega}{n}-1\right)^2\right)
 \left(\frac{T}{2 m_e c^2}\right)^n,
\end{equation}
where $R$ is the stellar radius, $\rho$ is the surface density of the electrons,
$\alpha$ is the fine structure constant, and non-dimensional frequency is
defined as

\begin{equation}
\label{j1}
 \tilde \omega = \frac{\omega}{\gamma \omega_H}
\end{equation}
The frequency $\tilde \omega$ of the $n$-th harmonic covers the interval
 $1/\gamma \le \tilde \omega \le 2 n$, so the relation
(\ref{f4}) is valid only in this frequency interval, and $Q_n$ is zero outside it.
Combining several harmonics together we obtain the final spectrum of the magneto-dipole
radiation.
The observed lines in the spectrum of
X0115+63\cite{b} are non-equidistant when they are interpreted as emission lines.
Besides, they are very broad as indicated in the Fig.\ref{f6} from \cite{b}.
It is shown in \cite{ba}, that magneto-dipole radiation of electrons
with very anisotropic momentum distribution could be the most natural explanation
of such spectrum. The calculated spectrum of magneto-dipole radiation corresponding to
the perpendicular electron temperature $20$ {ÊýÂ} is representer in Fig.\ref{f7}
from \cite{ba}. Note that the shape of this spectrum does not depend on
the parallel relativistic factor $\gamma_{\parallel}$, all frequencies are proportional
to the same factor $\gamma_{\parallel}$.
We need to know the parallel relativistic factor $\gamma_{\parallel}$ for
determination of the stellar magnetic field. This factor is depending on
the physical conditions in the accretion flow and may be different for
different X-ray pulsars.

\section{Magnetars}

Among more than 2000 cosmic gamma ray bursts (GRB) 4 recurrent sources
had been discovered, and were related to a separate class of GRB, called
soft gamma repeators (SGR). Besides observations of short, soft, faint
recurrent bursts, three of them had given giant bursts, most powerful among
all GRB. All four SGR are situated close or inside SNR, three of them show
long periodic pulsations. These properties had separated SGR, situated in our
or nearby galaxies, in a quite special class, very different from other
GRB, which are believed to have a cosmological origin at red shifts $z \sim 1$.
One SGR 1627-41 had been discovered by BATSE \cite{vkp99a} and
three other had been discovered in KONUS experiment in
1979 \cite{mg79,mg81,gil84}. Three of SGR show regular pulsations,
and for two of them $\dot P$ had been estimated, indicating to very
high values of the magnetic fields, up to $10^{15}$ Gs, and small age of
these objects. SGR have the following properties \cite{ffc99},
\cite{hur99a}-\cite{hur99e},
\cite{kou98,kou99,mg79},\cite{ma99a}-\cite{mc99c},\cite{mcs99,wkp99b}.

\medskip

\quad 1. SGR0526-66 \cite{mg79,mc99c}.

\medskip

It was discovered due to a giant burst of 5 March 1979, projected to the
edge of the SNR N49 in LMC. Accepting the distance 55 kpc to LMC, the peak
luminosity in the region $E_{\gamma}>30$ keV is equal to
$L_p=3.6\times 10^{45}$ ergs/s, the total energy
release in the peak $Q_p=1.6 \times 10^{44}$ ergs, in the subsequent tail
$Q_t=3.6 \times 10^{44}$ ergs. The short recurrent bursts have
peak luminosities in this region
$L_p^{rec}=3\times 10^{41}\,-\, 3 \times 10^{42}$ ergs/s,
and energy release $Q^{rec}=5\times 10^{40}\,-\, 7 \times 10^{42}$ ergs.
The tail was observed about 3 minutes and had regular pulsations with the
period $P\approx 8$ s. There was not a chance to measure $\dot P$ in this
object.

\medskip

\quad 2. SGR1900+14 \cite{ma99a,mc99c,kou99,wkp99b}.

\medskip

It was discovered first due to its recurrent bursts, the giant burst
was observed 27 August, 1998. The source lies close to the less than
$10^4$ year old SNR G42.8+0.6, situated at distance $\sim 10$ kpc.
Pulsations had been observed in the giant burst, as well as in the
X-ray emission observed in this source in quiescence by RXTE and
ASCA. $\dot P$ was measured, being strongly variable. Accepting the
distance 10 kpc, this source had in the region $E_{\gamma}>15$ keV:
$L_p > 3.7\times 10^{44}$ ergs/s, $Q_p > 6.8\times 10^{43}$ ergs,
$Q_t=5.2 \times 10^{43}$ ergs, $L_p^{rec}=2\times 10^{40}\,-\,
4\times 10^{41}$ ergs/s, $Q^{rec}=2\times 10^{39}\,-\, 6\times
10^{41}$ ergs, $P=5.16$ s, $\dot P=5 \times 10^{-11}\,-\, 1.5\times
10^{-10}$ s/s. This source was discovered at frequency 111 MHz as a
faint, $L_r^{max}=50$ mJy, radiopulsar \cite{shi99} with the same
$P$ and variable $\dot P$ good corresponding to X-ray and gamma-ray
observations. These values of $P$ and average $\dot P$ correspond to
the rate of a loss of rotational energy $\dot E_{rot}=3.5 \times
10^{34}$ ergs/s, and magnetic field $B=8 \times 10^{14}$ Gs. The age
of the pulsar estimated as $\tau_p=P/2\dot P=700$ years is much less
than the estimated age of the close nearby SNR. Note that the X-ray
luminosity of this object $L_x=2\times 10^{35}\,\, - \,\, 2\times
10^{36}$ ergs/s is much higher, than rate of a loss of rotational
energy, what means that rotation cannot be a source of energy in
these objects. It was suggested that the main source of energy comes
from a magnetic field annihilation, and such objects had been called
as magnetars \cite{dt92}.

\medskip

\quad 3. SGR1806-20 \cite{kou98,hur99e}.

\medskip

This source was observed only by recurrent bursts. Connection with
the Galactic radio SNR G10.0-03 was found. The source has a small but
significant displacement from that of the non-thermal core of this SNR.
The distance to SNR is estimated as 14.5 kpc. The X-ray source observed
by ASCA and RXTE in this object shows regular pulsations with a period
$P=7.47$ s, and average $\dot P=8.3\times 10^{-11}$ s/s. As in the previous
case, it leads to the pulsar age $\tau_p \sim 1500$ years, much smaller
that the age of SNR, estimated by $10^4$ years. These values of $P$ and $\dot
P$ correspond to $B=8\times 10^{14}$ Gs. $\dot P$ is not constant, uniform set
of observations by RXTE gave much smaller and less definite value
$\dot P=2.8(1.4)\times 10^{-11}$ s/s, the value in brackets gives 1$\sigma$
error. The peak luminosity in the burst reaches
$L_p^{rec}\sim 10^{41}$ ergs/s in
the region 25-60 keV, the X-ray luminosity in 2-10 keV band is
$L_x\approx 2\times 10^{35}$ ergs/s is also much higher than the rate
of the loss of rotational energy (for average $\dot P$) $\dot E_{rot}
\approx 10^{33}$ ergs/s.

\medskip

\quad 4. SRG1627-41 \cite{ma99a,vkp99a}.

\medskip

Here the giant burst was observed 18 June 1998, in addition to numerous
soft recurrent bursts. Its position coincides with the SNR G337.0-0.1,
assuming 5.8 kpc distance. Some
evidences was obtained for a possible periodicity of 6.7 s, but giant burst
does not show any periodic signal \cite{ma99a}, contrary to
two other giant burst in SGR. The following characteristics
had been observed with a time resolution 2 ms at photon energy
$E_{\gamma}> 15$ keV:
 $L_p \sim 8\times 10^{43}$ ergs/s,
$Q_p \sim 3\times 10^{42}$ ergs,
no tail of the giant burst had been observed.
$L_p^{rec}=4\times 10^{40}\,-\, 4\times 10^{41}$ ergs/s,
$Q^{rec}=10^{39}\,-\, 3\times 10^{40}$ ergs. Periodicity in this source is
not certain, so there is no $\dot P$.

To measure $\dot P$ the peaks of the beam are compared during long
period of time. Both SRG with measured $\dot P$ have highly variable beam
shapes, what implies systematic errors in the result. Another source of
systematic error comes from the barycenter correction of the arriving
signal in
the source with an essential error in  angular coordinates. This effect is
especially significant for determination of $\ddot P$ \cite{biskp93},
but when observational shifts are short the error in the
coordinates could not be extracted easily. Earth motion around the Sun, as
well as the satellite motion around the Earth may influence the results.
Nevertheless, independent measurements of $P$ and $\dot P$ in such
different spectral bands as radio and X-rays gave similar results for
SGR1900+14.

The physical connection between SGR and related SNR is not perfectly
established: SNR ages are much larger, than ages of SNR estimated
by $P$ and $\dot P$ measurements, and all four SGR are situated at the
very edge of the corresponding SNR, or well outside them. Using the pulsar
age estimation we come to conclusion of a very high speed of the neutron star
at several thousands km/s, exceeding strongly all measured speeds of
radiopulsars.
Physical properties of the pulsars observed in SGR are very unusual.
If the connection with SNR is real, and the distances to SGR are the same
then the pulsar luminosity
during giant bursts is much larger than the
critical Eddington luminosity. As follows from simple physical reasons,
when the radiation force is much larger than the force of gravity, the matter
would be expelled at large speed, forming strong outflow and dense envelope
around the neutron star which could screen pulsations. No
outflowing envelope around SGR have been found.
The  large difference between $\dot E_{rot}$ and average $L_x+L_{\gamma}$
luminosity needs to suggest a source of energy, much larger than the one
coming from the rotational losses. It is suggested that in "magnetars" the
energy comes from the magnetic field annihilation
\cite{dt92}. It is rather surprising to observe the
field annihilation without formation of relativistic
particles, where considerable part of the released energy should go.
Radio emitting nebula should be formed around the SGR, but
had not been found.  So, it is not possible to exclude that there is no
physical connection between SNR and SGR, that SGR are much closer objects,
their pulsar luminosity is less than the Eddington one, and their
magnetic fields are not so extremely high.

There is a striking similarity between SGR and special class of
X-ray pulsars called anomalous X-ray pulsars (AXP). Both have periods
in the interval 6 - 12 seconds, binarity was not found, spin-down of the
pulsar corresponding to small age $\sim 1000$ years, observed irregularities
in $\dot P$ (see e.g. \cite{mel99}). The main difference is an absence of
any visible bursts in AXP, characteristic for SGR. Suggestions had been
made about their common
origin as "magnetars" \cite{kou98,mel99}. Other
models of AXP had been discussed (see e.g. \cite{mis98}). We
may expect that establishing of the nature of AXP would help strongly
for the determination of the nature of SGR.

\section{Discussion}

Interpretation of SGR as "magnetars", and the hypothesis about
the magnetic field annihilation as the main source of energy arises
questions, when we compare this model with the well established
astronomical data. The giant explosions in GRB produce huge amount of
energy, $> 10^{44}$ ergs in gamma region, and comparable amount in
relativistic particles, generated during the field annihilation. 6 orders
of magnitude lower energy output in the form of relativistic particles,
produced during starquakes in the Crab pulsar make a visible changes
in the morphology of the Crab nebula \cite{sca}.
There are no data indicating the changes in the SNR structure after any giant
burst of GRB, while it should be possible to observe the results of such a
production of such a huge amount of energy in a short time. Magnetic field
annihilation is accompanied by a particle acceleration, so SGR should become
a bright radio source, at least after the giant burst. Nevertheless, SGRs are very
faint radio sources, and do not show increase of radio emission during  giant bursts.

Comparison of giant bursts in SGR and short GRB (duration less than 2 seconds)
shows a big resemblance of both
events. They have similar length, structure and spectra, which are harder than spectra
of long GRB (\cite{cmo99}).
If SGR would be about 10 times more distant objects we could see only their giant
bursts, which undoubtfully would be interpreted as short GRB.
The galaxies of our local group, containing SGR would be observed as sources of short
GRB. Having in mind, that Andromeda galaxy is about 3 times heavier than the Galaxy
\cite{allen}, we should expect about 10 SGR which could produce
a comparable number of giant bursts
during the observational time ( $\sim 30$ years), visible as short GRB
\cite{bis00,bis01,bis02}.
Analysis of BATSE and KONUS catalogs does not give any evidence about GRBs
coinciding  with any Local Group galaxy, including Andromeda \cite{mg81,cat4}
Better statistical estimation and search in BATSE/KONUS data are desirable for
investigation of connection between short GRB and SGR.

The present interpretation of SGR suggests a neutron star with very high magnetic field,
which is very short living, and should not be visible in its present view
after about 1000 years. During this time the neutron star will not have time to
cool enough, so it will be visible at least as soft X-ray source
with a temperature $>3 \cdot 10^5$ K during about $10^5$ years \cite{yak01}.
It would be interesting to look for such remnants in the existing and future
X-ray surveys. No candidates for such remnants are found until now.
In total we could expect about $10^8$ neutron stars as remnants of SGR
\cite{bis02}, and about 1000 of them should have surface temperatures exceeding
$>3 \cdot 10^5$ K.

Mechanism of "magnetar" emission in SGR suggests magnetic field annihilation as the
main source of the energy. We will not discuss the physics of such process, which looks
out very strangely, because it is unclear how it is possible to annihilate the
field, which is produced by electrical currents deep inside the neutron star, in its
core, or at least in its crust. Let us consider the results of such annihilation.
Solar flashes produced by field annihilation are
characterized by very broad spectrum, from radio to gamma.
 Is remains unclear how it is possible to provide $\gamma$-radiation by magnetic
field annihilation without appearance of large number of ultra-relativistic
particles and strong nonthermal emission in other spectral bands.

The estimation of the magnetic fields of "magnetars" in SGR is done using
the usual formula (ref{ref2}), which gives estimation of the magnetic fields in
rotationally powered pulsars. In the situation, when $L_{tot}\gg \dot E_{rot}$
the decrease of the rotational period may be connected not only with the
dipole-like radiation, leading to (ref{ref2}), but may be much more effective due to
ejection of relativistic particles (pulsar wind) induced by the field annihilation.
It will lead to much smaller values of the estimations of the SGR ages, which are
too small in comparison with the corresponding SNR ages even with this
artificial interpretation. As was indicated in \cite{mrl99} there are no
reasons to use (\ref{ref2}) with the measured $P$ and $\dot P$ for estimations
of SGR ages.

Contrary to SGR and AXP the age and magnetic field estimations of radiopulsars
do not have such contradictions and seems quite reliable. During last few years
several radiopulsars have been discovered, which have very similar periods and
magnetic fields, but otherwise behave like ordinary pulsars. The first discovery
of 8.5 s pulsar J2144-3933 with a normal magnetic field $B=2.0 \cdot 10^{12}$ Gs,
and large characteristic age $\tau=2.8 \cdot 10^8$ years
was reported in \cite{yom99}. The period of this pulsar is well inside the interval
6-12 s, characteristic for AXP, but magnetic field and age are similar to the
majority of radiopulsars. Farther discoveries revealed existence of radiopulsars
with very high fields, subscribed to "magnetars".
Pulsars PSRs J119-6127 and J1814-1744 have been discovered \cite{cam00}. These
pulsars with periods 0.407 s and 3.98 s, have magnetic fields
$B=4.1 \cdot 10^{13}$ Gs and $B=5.5 \cdot 10^{13}$ Gs, and characteristic ages
$\tau=1.6 \cdot 10^3$ years and $\tau=8.5 \cdot 10^4$ years respectively.
In these two pulsars only magnetic fields are unusually high but other
characteristics are normal, with the second period on the edge of the distribution.
In two subsequently discovered PSRs J1847-0130 and J1718-37184
magnetic fields are even higher, and all characteristics
are similar to the "magnetars". These
pulsars have periods 6.7 s and 3.4 s, magnetic fields
$B=0.94 \cdot 10^{14}$ Gs and $B=0.74 \cdot 10^{14}$ Gs, respectively,
and characteristic ages
$\tau=8.3 \cdot 10^4$ years is measured only for the first pulsar
\cite{mcl03,mcla03}. We may expect farther discoveries of radiopulsars
with large periods and high magnetic fields, like prescribed to "magnetars"
in AXP and SGR. These
discoveries create additional doubts to the "magnetar" model of SGR and AXP.

\section{Conclusions}

\begin{enumerate}
\item Magnetic fields of radiopulsars are in good correspondence with
theoretical estimations.
\item  RP and LMXB have small magnetic fields, which very probably
had been decreased by damping or screening during accretion stage.
\item  Contradiction between high $B_{cycl}$ and other
observational estimations of $B$ in the LMXB Her X-1 may be removed in
the model of relativistic dipole mechanism of the formation of a hard
spectral feature by strongly anisortopic relativistic electrons, leading to
conventional value of $B\approx 5\times 10^{10}$ Gs.
\item  Very high magnetic fields in magnetar model of SGR needs farther
confirmation and investigation.
\end{enumerate}

{\bf Acknowledgements}

This work was partially supported by RFBR grant 02-02-16900, and RAN programm
"Nonstationary processes in astrophysics".

 \medskip

\end{document}